\documentclass[fleqn,10pt]{wlscirep}

\usepackage{tikz}
\usepackage{makecell}%
\usepackage{graphicx}
\usepackage{epstopdf}
\usepackage{ogonek}

\usepackage[markup=default]{changes}  
\setaddedmarkup{{\color{orange}\uwave{#1}}}
\setdeletedmarkup{{\color{magenta}\sout{#1}}}
\definechangesauthor[color=orange]{KB} 
\definechangesauthor[color=blue]{FE} 

\title{Multilayer Aggregation with Statistical Validation: Application to Investor Networks}

\definecolor{cc686e9}{RGB}{198,134,233} 

\definecolor{c00caff}{RGB}{0,202,255} 
\definecolor{cFinancial}{RGB}{46,144,144} 
\definecolor{cNonProfit}{RGB}{255,92,129} 
\definecolor{cRestWorld}{RGB}{95,198,19} 

\definecolor{cea8615}{RGB}{234,134,21} 
\definecolor{cgreeen}{RGB}{0,0,255} 

\author[1,*]{Kęstutis Baltakys}
\author[1]{Juho Kanniainen}
\author[2,3]{Frank Emmert-Streib}
\affil[1]{Laboratory of Industrial and Information Management, Tampere University of Technology, Finland}
\affil[2]{Predictive Medicine and Data Analytics Lab, Department of Signal Processing, Tampere University of Technology, Finland}
\affil[3]{Institute of Biosciences and Medical Technology, Tampere, Finland}
\affil[*]{Corresponding author, kestutis.baltakys@tut.fi}

\begin{abstract}
Multilayer networks are attracting growing attention in many fields, including finance. In this paper, we develop a new tractable procedure for multilayer aggregation based on statistical validation, which we apply to investor networks. Moreover, we propose two other improvements to their analysis: transaction bootstrapping and investor categorization. The aggregation procedure can be used to integrate security-wise and time-wise information about investor trading networks, but it is not limited to finance. In fact, it can be used for different applications, such as gene, transportation, and social networks, were they inferred or observable. Additionally, in the investor network inference, we use transaction bootstrapping for better statistical validation. Investor categorization allows for constant size networks and having more observations for each node, which is important in the inference especially for less liquid securities. Furthermore, we observe that the window size used for averaging has a substantial effect on the number of inferred relationships. We apply this procedure by analyzing a unique data set of Finnish shareholders during the period 2004--2009. We find that households in the capital have high centrality in investor networks, which, under the theory of information channels in investor networks suggests that they are well-informed investors. 
\end{abstract}
\begin{document}

\flushbottom
\maketitle
%
%
\thispagestyle{empty}

\section*{Introduction}


Scientific literature on multilayer networks has recently started to gain more attention \cite{kivela2014multilayer, de2013mathematical, boccaletti2014structure}, having important applications in the financial area \cite{bargigli2015multiplex, musmeci2016multiplex}. Recent empirical evidence has triggered a disagreement over conventional tractable financial models\cite{battiston2016complexity}. In finance, complex network theory \cite{newman_2003,em_be3} has mainly been applied to gauge the systemic risk posed by interconnected banks \cite{cimini2015systemic,battiston2016complexity, barucca2016network, haldane2011systemic, cont2010network}, but recently, multiple network inference methods have been developed to investigate trading behavior\cite{tumminello2012identification,ozsoylev2013investor} and portfolios\cite{gualdi2016statistically} in investor network research.

An investor network is a representation of a real-world complex system where institutional and private investors indirectly interact with each other by trading or owning securities. In general, network science methods allow for analyzing and gaining a clearer understanding of the intricate relationships between the components of this system, and a key advantage of such an approach is that it allows for visualizing the resulting networks \cite{bastian2009gephi, jacomy2014forceatlas2}. However, estimating investor networks is not straightforward, as links between investors are not directly observable. Instead, a link represents the abstract distance of a pair of investors in terms of trading behavior or portfolios. Therefore, the analysis requires investor-level transaction or portfolio data and an appropriate statistical inference method for inferring such networks from the data. Even though complex network methods have begun attracting attention to investor-level data\cite{ranganathan2017dynamics}, many methodological challenges remain, several of which we aim to address in this paper. For our analysis, we use data from a large shareholder registry to investigate the trading networks of different investor categories.

First, the main challenge in investor trading networks is considering multiple securities leading to a multilayer network representation. What if we wanted a simple network representation, which would have statistically significant relationships over multiple securities? Ever-changing investor behavior poses difficulties for correctly inferring their relationships. Most likely, performing network inference for a whole period will not reveal the whole picture, as localized relationships between investor categories occurring at different periods might be diluted when we look at longer horizons. At the same time, static networks inferred over a whole period do not provide information on how node relationships evolve over time. In order to analyze the varying associations between investor categories, we use a simple, window-based analysis to recover the time-evolving networks of investor category interactions. Moreover, having a sequence of network snapshots, one might want to summarize the most important reoccurring relationships over the whole period. Therefore, we propose a multilayer aggregation approach that can address this challenge by integrating an ensemble of networks, resulting in a network that captures the most significant consistencies in investor relationships over multiple estimation periods and many securities. We also consider the influence of window size on the resulting aggregated networks\cite{emmert2010influence}. As we show, this approach allows for producing robust network structures using all of the transaction data over multiple securities and estimation periods without discarding a single transaction.

Second, we can think of investor trading as a data generation process that produces observations (transaction data) based on unobservable trading mechanisms. For example, trading algorithms have specific trading rules, and household investors with more or less intuitive trading strategies can have certain (stochastic) mechanisms, which are impossible to observe directly. The point is that the data set of observable transactions is just one realization of the underlying data generation process driven by certain mechanisms. Therefore, one might wonder which data sample to use for the network inference—all the transaction data together or one or more sub-samples of the full set of trading data. In addition, in our case, the investor category consists of many investors, and we want to prevent cases where a couple of active investors or investors who trade large volumes overshadow the behaviors of other investors in the category. In our approach, we address these problems by performing the lowest resolution bootstrapping at the investor transaction level. An empirical demonstration shows that the results clearly differ between the conventional approach of using the full data set directly and our data bootstrapping approach.

Third, the transaction data for network inference suffers from a high-dimension, low-sample size problem\cite{bernardo2003bayesian}, as the number of investors exceeds the number of trading days. Estimating investor networks based on trading patterns requires long observation periods and sufficient data for each investor \cite{tumminello2012identification}. Since the majority of household investors are rather inactive, only a fraction of investors—the active ones—can be included in the analysis. The exclusion of inactive investors leads to the description of a sub-system; therefore, the conclusions can be difficult to generalize at the market level. In this paper, we solve this problem by assigning investors to categories according to investor attributes that are available in the data set. Such a categorization allows us to reduce the number of variables in the system significantly, but we do not exclude data, as the categories contain aggregated data from the whole system. Importantly, this approach allows for considering inactive investors and less liquid stocks with fewer trading events. The size of such a categorized network remains the same over time, whereas the size of a network of individual investors can change over time, depending on the activeness of the investors. Since investor categories are based on real attributes, we can characterize the nature of each category.

We contribute to the methodological research literature by introducing three building blocks that can be used together or separately:  the investor categorization, transaction data bootstrapping and network aggregation. The main contribution of this paper is that we propose the use of a tractable multilayer and multistep aggregation procedure, by which we aggregate information from multiple layers using statistical validation. Methodologically, this approach can be used for different non-financial applications, with various network estimation methods, even for observable networks. Multilayer aggregation procedure can be used stand alone in cases when the underlying networks are either directly observable of the input is a network. In that case no data bootstrapping is necessary. 

In contrast to our paper, none of the existing procedures provide tractable procedure for the aggregation of binary network layers using statistical validation. The paper that mostly closely resembles ours regarding its topic, proposes an ensemble-based network aggregation \cite{zhong2014ensemble} method that leverages the rank-product method \cite{breitling2004rank} to improve the accuracy of gene network reconstruction. However, the algorithm is intended to integrate gene networks inferred using different methods and genomics data sets. Other trivial network ensemble aggregation procedures include maximum and mean rules\cite{polikar2006ensemble}. Another recent paper\cite{de2015structural} proposes a method for reducing the complexity of multilayer networks by aggregating the redundant layers while retaining the pertinent information about the whole system. In practice, the goal of their method is to combine similar layers and keep dissimilar layers apart. The objective of our research is different from Ref. \citen{de2015structural}; we are looking for the most important relationships that span multiple layers, rather than keeping information about different layers. The principal of our procedure is consistent with that of Ref. \citen{de2012bagging}; however, we are aggregating layers wheres Ref. \citen{de2012bagging} is for the inference of unlayered data.

Secondly, we contribute the literature on investor networks by introducing transaction bootstrapping, which improves the inference of the networks. Additionally, in the investor network inference, we consider investor categories, instead of individual investors. This makes the system interpretable in economic and sociological terms. As a fourth building block, we use already existing network inference methods to identify links between investors. In addition to the methodological contribution, this is the first paper that provides investor network graphs over multiple securities and time-windows. We provide empirical evidence that households in Helsinki are the most central investor category that, under the theory of information channels in investor networks\cite{ozsoylev2013investor}, this category is shown to represent well-informed investors.

Overall, our framework can be summarized as follows (in chronological order):
\begin{itemize}
	\item [--] \textbf{Investor categorization}, where all investors in the analyzed data set are assigned to only $99$ categories based on their economic and social attributes. Categorization allows to keep the number of nodes in the network constant, with sufficient transaction data for each one of them. \textit{(Investors $\rightarrow$ Investor Categories})
	\item [--] \textbf{Transaction data bootstrapping}, where the analyzed data is uniformly resampled into multiple data sets for better statistical validation in the network inference. The advantage of the bootstrap is that it does not require any assumptions about the data distribution and it addresses the issue of finite observations. \textit{(Dataset $\rightarrow$ Resampled Datasets)}
	\item [--] \textbf{Network inference}, where a selected inference method is used to identify edges between investor categories for each resampled data set to produce {an ensemble} of networks. In this regard, we use two existing methods: Conservative Causal Core Network (C3NET) \cite{altay2010inferring} and Minimum Spanning Trees (MST). In fact, \emph{any network inference method} \cite{mantegna1999hierarchical, tumminello2005tool, peng2009partial, boginski2005statistical, onnela2004clustering} that produces or can be converted into binary, non-weighted networks could be applied.  \textit{(Dataset $\rightarrow$ Network)}
	\item [--] \textbf{Network aggregation}, where by using statistical validation, a network ensemble is aggregated to identify significant relationships that appear across the set of networks. In our demonstration we employ the aggregation procedures in three different ways revealing the most important relationships: during bootstrapping , where relationships arise for the same security during the same time period; in the time-wise aggregation, where the relationships are observed for the same security in different time periods; and in the security-wise aggregation, where the relationships are observed in the same time period over multiple securities. \textit{(Network Ensemble $\rightarrow$ Aggregated Network)}
\end{itemize}

To demonstrate our multilevel aggregation approach, we use an investor-level transaction data set obtained from Euroclear Finland Ltd for our analysis. It includes transactions from 2004-01-01 to 2009-12-31 of all domestic investors that traded stocks listed on the Nasdaq OMX Helsinki Exchange. Each transaction also contains meta-data about the investors (the same data set is used, for example, in refs. \citen{tumminello2012identification,grinblatt2000investment,berkman2014informed}, while ref. \citen{ozsoylev2013investor} uses a similar data set of trades on the Istanbul Stock Exchange). In this data set, the attributes used to categorize investors include gender, year of birth, and postal code for households and sector code for institutions. These attributes allow us to define $99$ investor categories on which our analyses are performed.

\begin{figure}[!hbtp]
\centering
\includegraphics[scale=0.8]{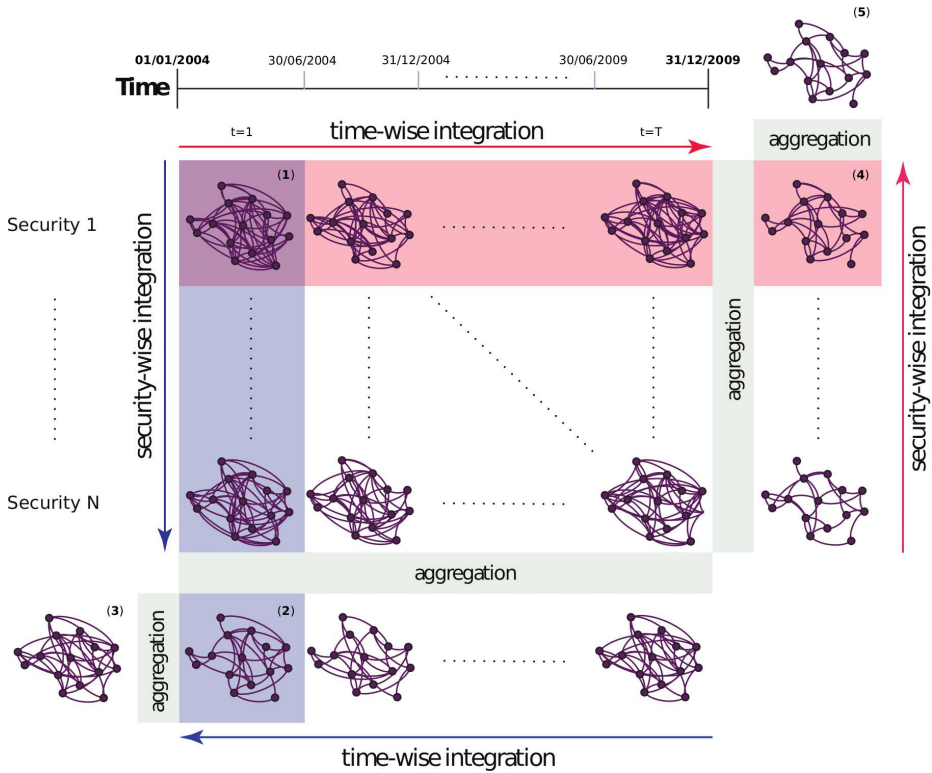}
\caption{\label{BaggingLayers}\textbf{\emph{Aggregation of ensemble of individual networks}}. A set of networks in the main matrix containing two information layers—time and securities (networks adjacent to network (1)). Each of these networks is a result of applying a chosen network inference algorithm for the corresponding transaction data sets. Two information integration approaches are possible: one-layer integration of either securities (2) or time (4), for each period and security, respectively, or a multilayer approach, where both information layers are fully integrated, leading to networks (3) and (5).}
\end{figure}

Before providing results on our data set, let us elaborate the main principle of aggregation, while leaving the technical details of statistical validation into Methods section. Fig. \ref{BaggingLayers} demonstrates how the ensemble of individual networks are aggregated. Overall, we have two different layers in investor networks. The first layer indicates securities and the second one indicates time. Interestingly, there are two different ways to integrate over these variables, indicated by the blue and red arrows. We show in the following that the results highlight different characteristics of the data.

For each time step and each security, we want to extract a network. These networks cannot be directly observed, but they are estimated using the transaction data set. By bootstrapping this data set, we generate $B$ bootstrap data sets. Network inference is applied to each of these $B$ data sets, resulting in an ensemble of $B$ networks. The aggregation of the $B$ networks results in \emph{one network}, indicated by (1) (see Figure \ref{BaggingLayers}). Adjacent networks in the main matrix are similarly inferred for other time steps and securities. Each column is an ensemble of networks that contains information about trading relationships for different securities during the same time step, while each row is a network ensemble that contains information about trading relationships in individual securities over different time steps. In the following, we first describe the security-wise integration and then the time-wise integration.

Initially, we integrate the security-wise information for each time step contained in the columns. Network (2) represents an aggregated network for time step 1 over all securities. Repeating a similar analysis for each of the $T$ different time steps results in further networks for the corresponding cases. To combine these $T$ networks, we perform aggregation again, resulting in one final network, indicated by (3) in the figure. The blue arrows in the figure represent the aforementioned steps. Alternatively, one can perform a time-wise integration first in a similar way. This type of integration follows the red arrows in the figure and applies 2 times the aggregation method because two integration steps are required. This leads to the final network indicated by (5). Interestingly, even though the final networks (3) and (5) summarize the same information, because of the different aggregation order, the captured relationships might be different, as shown in the results section.

In the next section, we present the results from the method application to our data set. We begin by applying our proposed techniques to single security networks. We investigate the impact of transaction bootstrapping on the network inference problem and compare a network inferred over the whole period to an aggregated network from a set of network snapshots. Next, we investigate multiple security networks. First, we use the aggregation technique to summarize information about trading in multiple securities and then we perform a two-layer aggregation, summarizing the information given by a series of network snapshots for a set of securities.

\section*{Results}
In this section, we describe the network inference and aggregation process over single and multiple securities by performing the analysis over the whole period of analysis and multiple non-overlapping sub-periods. Mutual information (MI) values are estimated from daily net volume time series for each investor group pair. Any methods that produce or can be converted into binary, non-weighted networks can be used and in this paper we employ two existing algorithms. As a main algorithm, we employ \emph{Conservative Causal Core} (C3NET)  (see Methods for details) for network inference from the MI estimates in order to demonstrate our aggregation method. Moreover, we compare the results for single security trading network obtained using C3NET and the \emph{minimum spanning tree} (MST).

\subsection*{Single Security Networks}
We investigate single security networks using data on the most liquid security in the Helsinki stock exchange---namely, Nokia. In this section, we aim to demonstrate transaction bootstrapping impact on network inference and the aggregation of multiple networks inferred from different time periods. Also, we compare results obtained using two inference algorithms, C3NET and MST.

\subsubsection*{Network inference}

We begin our results section by comparing inferred networks using C3NET algorithm with and without transaction bootstrapping. By definition, C3NET allows for establishing as many links as there are nodes in the network, if each investor group has at least one statistically significant MI estimate with some other group. In our data set for Nokia from 2004-01-01 to 2009-12-31, using C3NET, we infer 90 links. Interestingly, even after completing the categorization, some investor categories do not have a sufficient number of Nokia transactions to estimate relationships. For the bootstrapped version of network inference, we perform 100 transaction sampling iterations and form a network for each of them using the C3NET algorithm.
The resulting ensemble of 100 networks contains $8\,853$  links with $1\,195$ different relationships. As a statistical null model for our ensembles, we choose the canonical Erd\H{o}s–R\'enyi  $G(n, p)$ model, with a fixed number of nodes and an ensemble probability of a random link (see the methods section for more details). A fully connected ensemble would have $n\times (n-1)/2\times B = 4\,851\times 100$ links; therefore, the probability of having a random link in the ensemble is estimated to be $p=8\,853/485\,100=1.82\times 10^{-2}$. By choosing a significance of $\alpha=0.01$ and adjusting it by the number of tests we perform ($1\,195$), we conclude, that a relationship must be observed in at least $10$ networks for it to be considered non-randomly occurring. The bootstrapped version identifies a total of $197$ relationships that are statistically significant. Hence, the topology is no longer limited to one link per node. Almost all relationships from the non-sampled C3NET network are found also in the bootstrapped version—that is, $77$ out of $90$.

\begin{figure}[hbtp]
\centering
\textbf{Without transaction bootstrapping} \hspace{2cm} \textbf{With transaction bootstrapping}\\

\vspace{0.5cm}
\rotatebox{90}{\textbf{Aggregated 6 month window networks} \hspace{2cm} \textbf{Whole period networks}}
\hspace{10px}
\includegraphics[width=0.8\textwidth]{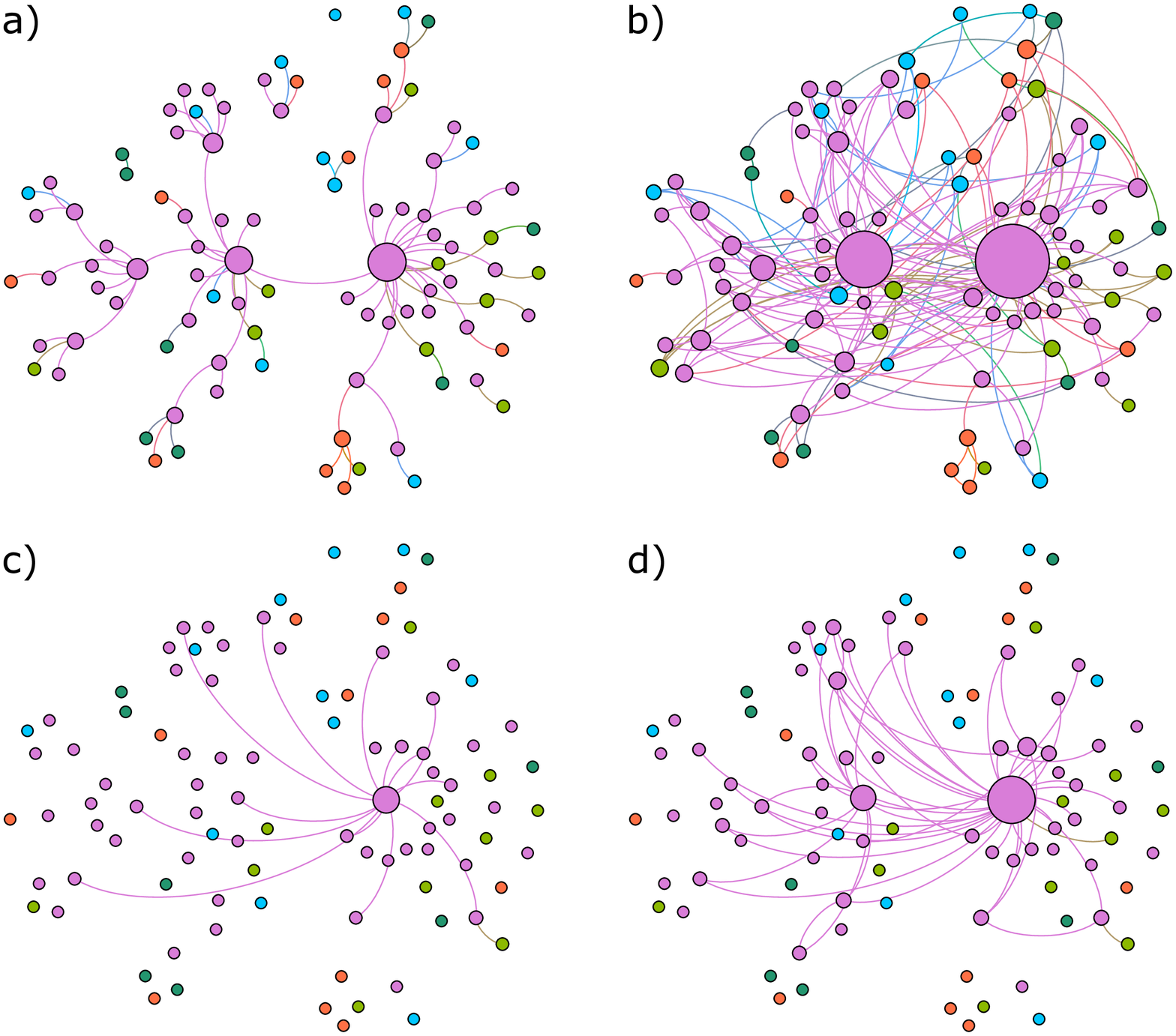}
%
%
%

\caption{\label{Nokia} Four networks of investor group trading relationships in Nokia security. Investor group positions are fixed in all four plots. Node sizes depend on node degrees in each network. The first network (a) is inferred using the C3NET algorithm on the original data set. The second network (b) is inferred by bagging C3NET. For the third (c) and fourth (d) networks, the whole six-year period is divided into 12 six-month  sub-periods. For each of those 12 sub-periods, a C3NET and bagged C3NET networks are inferred. Then those 12 networks are aggregated into a final network that covers the whole 6-year period.
\protect\tikz{\protect\draw [black, thin, fill=cc686e9] (4pt,4pt) circle [radius=4pt];} - Households, 
\protect\tikz{\protect\draw [black, thin, fill=cea8615] (4pt,4pt) circle [radius=4pt];} - Non-profit organizations,
\protect\tikz{\protect\draw [black, thin, fill=cRestWorld] (4pt,4pt) circle [radius=4pt];} - Other companies
\protect\tikz{\protect\draw [black, thin, fill=c00caff] (4pt,4pt) circle [radius=4pt];} -  Financial and insurance companies, 
\protect\tikz{\protect\draw [black, thin, fill=cFinancial] (4pt,4pt) circle [radius=4pt];}
 - Government institutions.}
\end{figure}

The two networks are depicted in (a) and (b) sub-plots of Figure \ref{Nokia}. Both networks identify the same nodes as most connected, and the four most connected nodes represent households. Specifically, the most connected node represents mature Helsinki households, followed by the same age group of western Tavastians, then middle-aged western Tavastians, and finally, mature northern Finnish households. The most connected non-household groups in the bootstrapped version are non-profit organization from eastern Tavastia, non-financial company from Ostrobothnia, and financial insurance company from northern Savonia, with six relationships each.

\begin{figure}[hbtp]
\centering
\includegraphics[width=\textwidth]{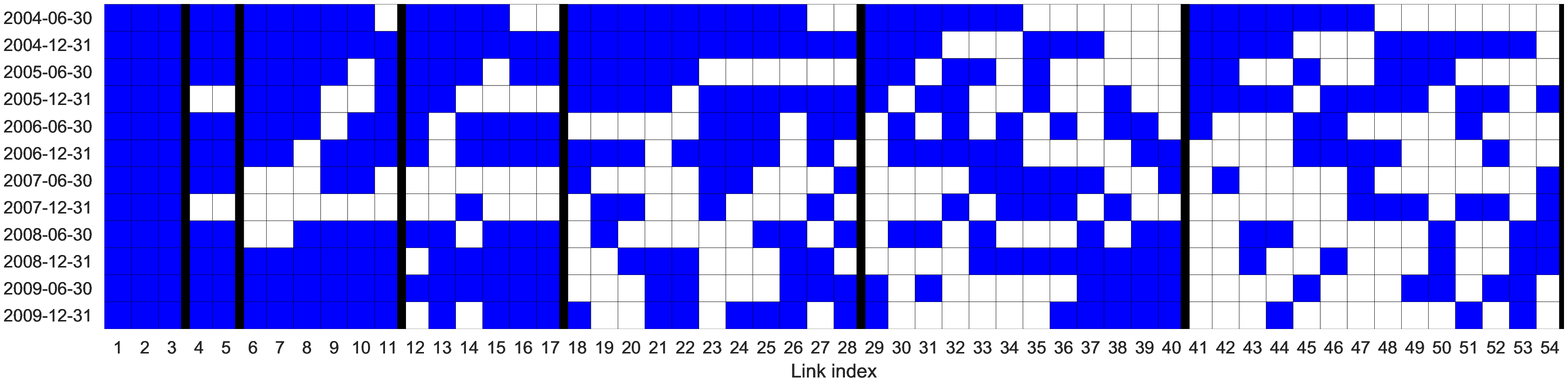}
\caption{\label{Nokia6monthSignLinks}54 most re-occurring links in 12 Nokia networks estimated over non-overlapping 6-month periods.
\protect\tikz{\protect\draw [black, thin, fill=cgreeen] (0pt,0pt) rectangle (8pt, 8pt);} - inferred relationship, 
\protect\tikz{\protect\draw [black, thin, fill=white] (0pt,0pt) rectangle (8pt, 8pt);} - no relationship.}
\end{figure}

\subsubsection*{Time-wise network aggregation}

The third and fourth networks for Nokia security in Figures \ref{Nokia} (c) and (d) are obtained by aggregating two 12-network ensembles inferred from non-overlapping 6-month periods covering the whole 6-year period analyzed. As in the previous section, we compare Nokia networks inferred with and without transaction bootstrapping. The number of relationships in the non-bootstrapped version ensemble varies from $63$ to $80$ and from $182$ to $216$ in the bootstrapped version. A total of $1\,420$ different relationships are observed throughout the 12 networks in the transaction bootstrapped network ensemble and the total number of links in the ensemble is $2\,361$, while in the non-bootstrapped version, the numbers of relationships and links are $673$ and $858$, respectively. Each network contains 99 nodes, and therefore, the total possible number of links in the ensemble is equal to $12\times 4\,851 = 58\, 212$ and the probabilities of having a random link are estimated to be $p=2\,361/58\,212\approx 4.05\times 10^{-2}$ and $p=858/58\,212 \approx 1.47\times 10^{-2}$.
Again, by choosing the statistical significance of $\alpha=0.01$ and adjusting for the number of tests performed, a link must appear at least $5$ times in order to be aggregated into the final network for the bootstrapped version and $4$ times for the non-bootstrapped version. From Table \ref{tab:NokiaLinkOcc}, we see that in the bootstrapped version, $54$ links appear at least $5$ times in the $12$ networks, and Figure \ref{Nokia6monthSignLinks} shows the link occurrence in the ensemble. In the latter figure, we can see that some relationships are accumulated in consecutive periods while others are more scattered over time.
\begin{table}[hbtp]
\centering
\begin{tabular}{lrrrrrrr|r|rrrr}
\toprule
Number of occurrences &  12 &  11 &  10 &  9  &  8  &  7  &  6  &  \textbf{5}  &   4  &   3  &   2  &    1  \\
\midrule
Links      &   3 & 0&  2 &   6 &   6 &  11 &  12 &  14 &  38 &   96 &  312 &   920 \\
Cumulative &   3 & 3&  5 &  11 &  17 &  28 &  40 &  54 &  92 &  188 &  500 &  1420 \\
\bottomrule
\end{tabular}
\caption{\label{tab:NokiaLinkOcc} Number of link occurrences in the Nokia ensemble inferred over non-overlapping six-month periods using the bootstrapped version of C3NET. We can see that only one link appears in all 12 networks, while 992 links appear only once.}
\end{table}

\begin{figure}
\centering
\includegraphics[width=0.6\textwidth]{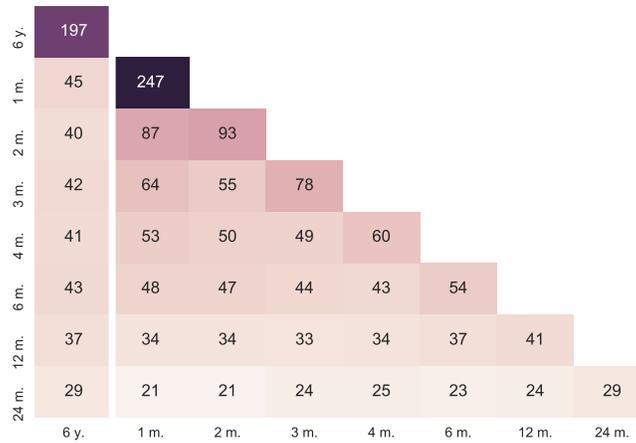} 
\caption{\label{Nokia_overlaps} Link overlap in variously inferred networks for Nokia. 6y. is the network inferred using transaction bootstrapping over the whole period. All the other networks are inferred on shorter windows (1, 2, 3, 4, 6, 12, 24 months) and then aggregated into a network that covers the whole period under analysis. From the figure, we can observe that most of the relationships inferred over the longer observation windows are also found in the shorter window analyses.}
\end{figure}

From Figure \ref{Nokia_overlaps} we can see that $43$ links overlap with the bootstrapped version of C3NET for the whole period under analysis. Further, for the non-bootstrapped version, 14 relationships are inferred after time-wise aggregation. Of those 14 relationships, 14 also appear in the bootstrapped version. All nodes, but two non-financial investor groups that have relationships, are households. A visual inspection of all four networks in Figure \ref{Nokia} reveals that the most important set of nodes in both networks inferred from the whole transaction data set is also identified as central in networks aggregated from various time window analyses.

\subsubsection*{Choice of network inference method}
The multilayer aggregation framework works independently of the choice for network inference, because it is applied to an ensemble of networks and not the underlying data. Here we present Table \ref{MSTvsC3NET} to compare the obtained aggregated Nokia networks using the C3NET and MST network inference algorithms. In the table we can see that the two similarly strict network inference algorithms yield highly similar results. We can also see that once we use network aggregation, the resulting aggregated network does not comply the definitions of respective algorithms. For example, in the aggregated networks MST no longer has $(n-1)$ links,  where $n$ is the number of nodes in the network (here $n = 95$). Also, C3NET is no longer limited to at most $n$ links.

\begin{table}
\centering
\begin{tabular}{l|c|cccccccc}
{} &  \textbf{Without} & \multicolumn{8}{c}{}\\
{} &  \textbf{transaction} & \multicolumn{8}{c}{\textbf{With transaction bootstrapping }}\\
{} &  \textbf{bootstrapping} & \multicolumn{8}{c}{ }\\
\midrule
{Length of (sub-)periods} &  \multicolumn{1}{c|}{6 y.} &  6 y. &  1 m. &  2 m. &  3 m. &  4 m. &  6 m. &  12 m. &  24 m. \\
\midrule
\multicolumn{10}{c}{Panel A: Links}\\
\midrule
MST              &   94 &   215 &   333 &   137 &    97 &    84 &    59 &     39 &     28 \\
C3NET            &   90 &   197 &   247 &    93 &    78 &    60 &    54 &     41 &     29 \\
\midrule
MST $\cap$ C3NET &   90 &   177 &   219 &    83 &    67 &    53 &    49 &     32 &     25 \\
\midrule
\multicolumn{10}{c}{Panel B: Nodes}\\
\midrule
MST              &   95 &    96 &    51 &    42 &    38 &    40 &    38 &     35 &     28 \\
C3NET            &   95 &    96 &    48 &    41 &    36 &    33 &    37 &     36 &     29 \\
\midrule
MST $\cap$ C3NET &   95 &    96 &    48 &    37 &    32 &    32 &    36 &     33 &     26 \\
\bottomrule
\end{tabular}
\caption{\label{MSTvsC3NET} Here we compare the number of estimated links and nodes having a link in the aggregated networks for Nokia security. As the length of time period, "6 y." is the network inference algorithms applied to the whole 6 year period. "1 m.", "2 m.", ... are the bootstrapped versions of network inference algorithms applied over 1, 2, ... month periods and then aggregated over all respective periods to cover the whole 6 years.}
\end{table}

\subsection*{Multiple Security Networks}

In the extant literature, investor trading networks have estimated for single securities (see Ref. \citen{tumminello2012identification}). In this paper, we show how to aggregate the networks of 100 securities into one. The security-specific networks are inferred using the whole 6-year period with transaction bootstrapping. Next, we examine the aggregation of networks with respect to securities and time-periods. Particularly, we explore how different the resulting aggregated networks are if we first integrate security-wise and then time-wise information versus in other way around. For simplicity, we present results for the bootstrapped version of C3NET  and results for MST are available upon request. 

\subsubsection*{Security-wise aggregation}
Here, we aim to incorporate information about investor group trading relationships in 100 securities over the whole 6-year period.
The number of inferred relationships across different securities ranges from $88$ to $261$, while the total number of detected relationships in the ensemble is $3\,218$. 
Subsequently, for the ensemble of 100 security networks, we apply the same aggregation procedure as before. 
From the observed number of links in the ensemble and total possible number of links in a fully connected ensemble of this size,
we estimate the probability of random links to be $p=20\,229/485\,100=4.17\times 10^{-2}$. Then, for a significance level of $\alpha=0.01$, 
we apply Bonferroni adjustment in $3\,218$ tests and end up with a threshold of $16$ link occurrences in the ensemble, which leaves $236$ links in the aggregated network. 
Households represent the majority of groups with relationships over multiple securities. Furthermore, two of the most central nodes are mature and middle-aged household investor groups from Helsinki, with 49 and 30 relationships, respectively. The two most central non-household investor groups are financial and non-financial companies in Helsinki, both with 9 relationships to other investor groups.

\subsubsection*{Two level aggregation}

\begin{figure}[hbtp]
\centering
\textbf{Security-wise $\Rightarrow$ Time-wise $\Rightarrow$ (3)} \hspace{5cm}\textbf{Time-wise $\Rightarrow$ Security-wise $\Rightarrow$ (5)}\\
\includegraphics[width=1\textwidth]{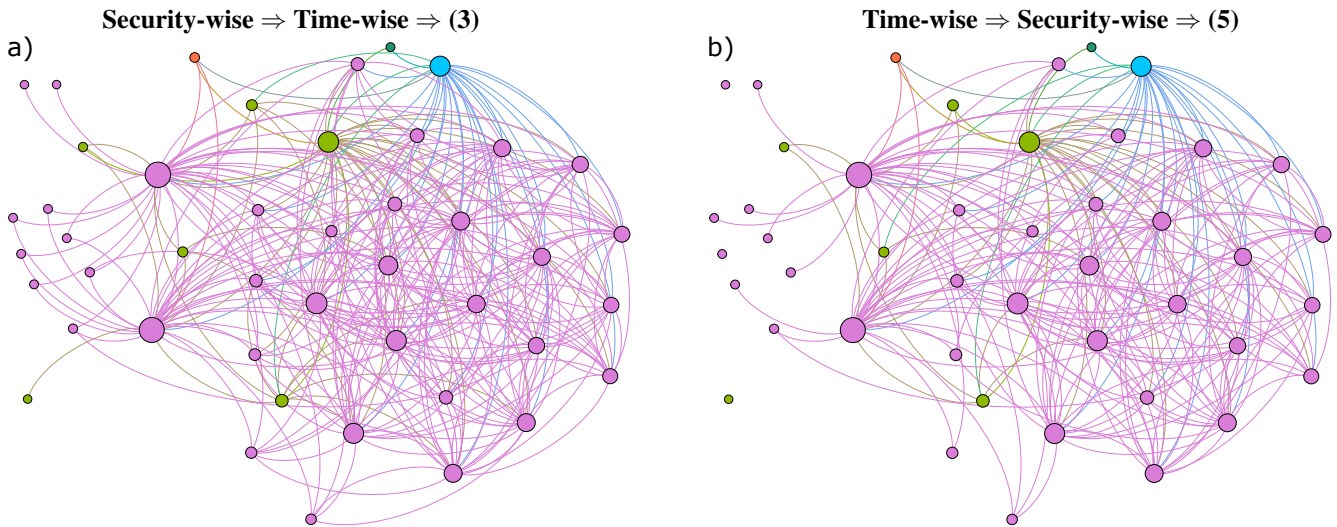}

\caption{\label{MultiNetworks}Networks summarizing investor group trading similarities in 100 securities over 6 years. The starting point for both networks is a set of $1\,200$ networks inferred for each security over 12 six-month, non-overlapping periods. The difference between networks comes from the aggregation order. The first network is first aggregated security-wise and then time-wise, while the third network is aggregated in reverse order. Network (3) in Figure \ref{BaggingLayers} represents network (a), while network (5) in the same figure represents network (b).
\protect\tikz{\protect\draw [black, thin, fill=cc686e9] (4pt,4pt) circle [radius=4pt];} - Households, 
\protect\tikz{\protect\draw [black, thin, fill=cea8615] (4pt,4pt) circle [radius=4pt];} - Non-profit organizations,
\protect\tikz{\protect\draw [black, thin, fill=cRestWorld] (4pt,4pt) circle [radius=4pt];} - Other companies
\protect\tikz{\protect\draw [black, thin, fill=c00caff] (4pt,4pt) circle [radius=4pt];} -  Financial and insurance companies, 
\protect\tikz{\protect\draw [black, thin, fill=cFinancial] (4pt,4pt) circle [radius=4pt];}
 - Government institutions.}
\end{figure}

\begin{figure}[!hbtp]
\centering
\includegraphics[width=\textwidth]{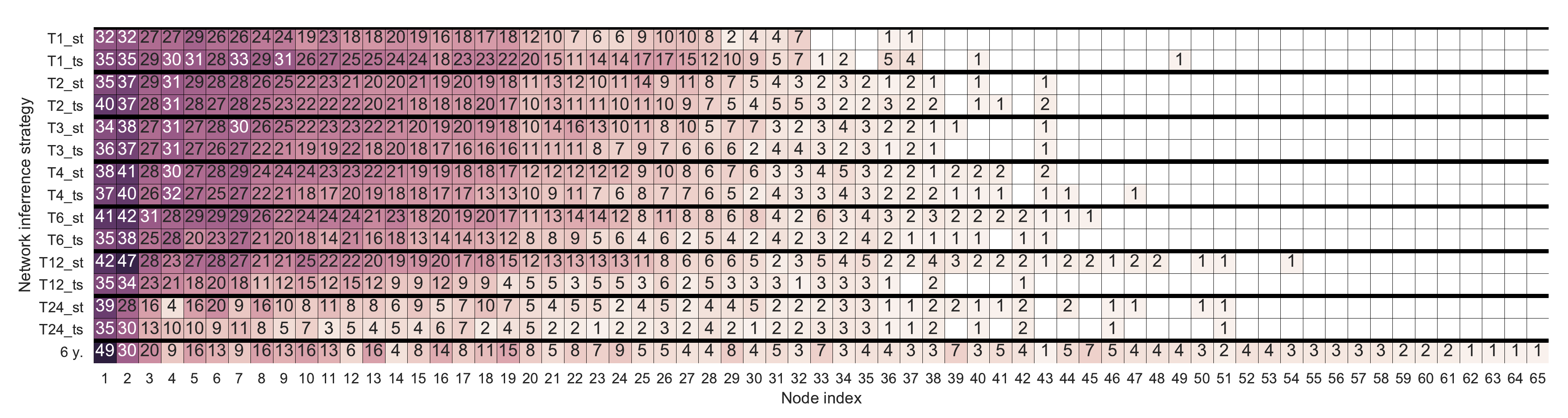}
\caption{\label{MultiSecTimeDegreeMap}Node degree comparison in final networks aggregated over 100 securities and non-overlapping time windows covering the whole 6-year period. $T\{W\}\_\{O\}$, where $W$ stands for the window size and $O$ stands for aggregation order, either security-wise or time-wise first. The row with \emph{6 y.} inference strategy stands for the network aggregated from networks that were inferred using the whole six-year data set for each security.}
\end{figure}

Here we leverage the previously introduced time-wise and security-wise network aggregation procedures. Our goal is to produce a single network that can summarize the trading relationship information inferred for 100 securities over multiple and various sizes time windows. We investigate networks inferred over seven different non-overlapping time windows—that is, 1, 2, 3, 4, 6, 12, and 24 months. Each security respectively has 72, 36, 24, 18, 12, 6, and 3 such networks, covering the whole 6-year period under analysis. Our starting point is a set of network ensembles inferred using bootstrapped C3NET algorithm for 100 securities for all analyzed time window sizes. For instance, in the case of the 6-month window, we have 12 networks for each of the 100 securities—that is, an ensemble of $12\times 100=1\,200$ networks (corresponding to the networks in the main matrix of Figure \ref{BaggingLayers}). We must also keep in mind that the aggregated network will differ depending on the order of information aggregation—that is, if relationship time-wise or security-wise information is summarized first. Accordingly, we describe the results of using both approaches and compare the final results. By performing the time-wise aggregation first, we end up with a 100-network ensemble, with one network for each security. Links in each network represent the most important reoccurring relationships in corresponding securities. Conversely, if we start with security-wise aggregation, we end up with an ensemble of 12 networks. Each of the 12 networks contains the most important relationships that are present over multiple securities, but this might be a different set of securities in each period. Next, for the two ensembles stemming from the first aggregation procedure, we perform the final aggregation, yielding a network summarizing the relationships of investor groups in their trading behavior over 100 securities for the whole period under analysis. However, the two final networks are not the same (see the networks in Figure \ref{MultiNetworks}). Table \ref{tab:FinalOverlaps} compares the links and nodes in the final networks for various window sizes. For each of the seven time windows, we obtain two networks, depending on the order of the aggregation procedure; thus, together with the security-wise aggregated network for the whole period from the previous section, we compare 15 networks. Figure \ref{MultiSecTimeDegreeMap} summarizes the node degrees in all 15 final networks.
Node degree sequences are highly correlated, with Spearman’s correlation ranging from $0.65$ to $0.99$.
Similar to the whole period security-wise aggregated network, networks in Figure \ref{MultiNetworks} identify mature and middle-aged household investor groups from Helsinki as the most central groups, while financial and non-financial company investor groups from Helsinki are most central non-household investor groups.

\begin{table}[ht]
\centering\begin{tabular}{c|rrrr|rrrr}
\toprule
Window &  \multicolumn{4}{c|}{Nodes}  &    \multicolumn{4}{c}{Links} \\
size & $ST\setminus TS$ &  $ST\cap TS$ &  $TS\setminus ST$ &  Jaccard &  $ST\setminus TS$ &  $ST\cap TS$ &  $TS\setminus ST$ & Jaccard \\
\midrule
1  &   0 &       34 &   4 &  0.8947 &    1 &    264 &  85 &  0.7543 \\
2  &   0 &       40 &   1 &  0.9756 &   19 &    282 &  15 &  0.8924 \\
3  &   1 &       39 &   0 &  0.9750 &   40 &    263 &   7 &  0.8484 \\
4  &   0 &       42 &   2 &  0.9545 &   46 &    259 &   8 &  0.8275 \\
6  &   3 &       42 &   0 &  0.9333 &   86 &    233 &   3 &  0.7236 \\
12 &  13 &       38 &   0 &  0.7451 &  132 &    176 &   4 &  0.5641 \\
24 &   5 &       42 &   0 &  0.8936 &   62 &     93 &  20 &  0.5314 \\
\bottomrule
\end{tabular}
\caption{\label{tab:FinalOverlaps}Summary of node and link overlap in various window size final networks. ST stands for the network where the first aggregation layer is security-wise (network (3) in Figure \ref{BaggingLayers}) and TS is the network where the first aggregated layer is time-wise (network (5) in Figure \ref{BaggingLayers}).}
\end{table} 

\subsection*{Financial interpretations}

The networks are designed to reflect associations in investors’ trading behavior and hence the high centrality reflects that there are many other groups of investors that behave in a mutually predictable way. Conversely, if an investor category is measured to have low centrality, then this investors in category are trading differently compared to other investor categories. From the point of view of information networks in financial markets \cite{ozsoylev2013investor}, high centrality of a node indicates that investors in the node have many private information channels from other investors. Therefore, given that association between investors trading patterns reflects information transfer, high centrality reflects good access to private information. In this paper we consider both positive and negative linkages to represent information transfer, that is two investors can employ the mutual information channel to trade in the same or opposite directions. Technically, we have used three centrality measures: Degree (fraction of nodes it is connected to), load centrality (fraction of shortest paths that include the given node), and closeness (1 divided by the average shortest-path distance for a given node).

Regarding the networks of investors over all the securities and 12 half-year periods (see Figure \ref{MultiNetworks}, the two most central investor categories are Middle-Age and Mature Household investors located in Helsinki. This result is robust across all the measures, i.e. degree, load centrality and closeness. A deeper investigation of the results shows household traders trade in a rather mutually predictable way between each other, though links to financial and non-financial companies exist, too. From the point of view of information propagation, these investors, i.e. households in the capital, are well-connected to other investor groups across the country. Potentially, this can be explained by rather strong and long-lasting internal migration and industrial changes in Finland to urban centres and especially to the capital of Helsinki (see, for example, ref. \citen{olli2001internal}). Particularly, such work-related migration can strengthen the social ties between different regions of the country. In addition, top-10 central nodes include not only household investors of different age groups and regions, but also Financial and Insurance Companies and Non-Financial Companies, which are among 3-7 most central nodes (the rank depends on the centrality measure). This can be seen also from Figure 5, where the node of Financial and Insurance Companies is with blue color and Non-Financial Companies in green. This means that the investment behavior of financial institutions is not very different from the behavior of other investor groups. In fact, this result does not support some earlier findings that institutional investors and households (all households collected together) exhibit different trading strategies (see ref. \citen{grinblatt2000investment}). Rather, based on our findings, there are large groups of households whose trading patterns are strongly associated with financial institutions, which we can observe from validated networks, aggregated over multiple stocks.

\section*{Discussion}

The advantage of our aggregation approach is that no arbitrary link-filtering threshold is needed. Instead, the algorithm adjusts this itself depending on a chosen significance level and the properties of the investigated network ensemble. We found that networks aggregated over multiple time-periods and inferred over the whole period significantly differed in the number of relationships inferred and the number of nodes having relationships. However, a similar set of nodes was identified as central in both cases.  Second, when the networks are aggregated over multiple securities and time-windows, two-layer aggregation yields different network descriptions depending on the order of information aggregation. It is worth mentioning that the aggregation of time-wise and security-wise trading relationships could be performed in a single step, in which case there would be no confusion about the aggregation order. However, in that case, the meaning of network relationships would be obscure. We would be neither certain that investor categories were similarly trading over a significant number of the same securities nor that they were trading similarly over a significantly large number of the same periods; further, the definition of a single step aggregation would be somewhere in-between, in some cases perhaps failing to meet both criteria. This question will be addressed in the future research.

Second, to the best of our knowledge, we are the first to propose the use of lowest resolution---that is, transaction-level---bootstrapping as the means for statistically validating investor network relationships. The advantage of transaction bootstrapping is that it enables network inference over shorter time windows to provide insight into the dynamics of these relationships. Most of the research has been focused on inferring static or time-invariant investor networks, and much less has been done to infer the dynamic relationships that are constantly evolving over time. Indeed, over the course of time, multiple interchanging processes may determine the behavior of investor categories, and such processes can be dynamic and stochastic. Therefore, investor behavior at each time point is dependent on these processes, and investor networks can undergo significant topological changes, rather than being invariant over time. Transaction bootstrapping is a viable strategy for network inference because it not only allows for assigning statistical significance to link existence but also enhances the robustness of the relationships to specific realizations of the trading outcome.

Finally, we introduced investor grouping into categories based on their attributes. This approach allows for performing any analysis by discarding less information. Also investor category networks based on investor attributes have not been investigated previously in the literature. The vulnerability of the investor categorization approach is that the ensuing analysis is ultimately dependent on the category definition. 

In the results section, we observed that Helsinki households represented the most connected investor category. Under the theory of information channels in investor networks\cite{ozsoylev2013investor}, this category is shown to represent well-informed investors. In fact, the important role of household investors has been identified in the literature\cite{grinblatt2001distance, grinblatt2001makes, kaniel2012individual, kaniel2008individual}. For example, according to \citen{kaniel2008individual}, households are contrarian traders, leading them to serve as liquidity providers to institutional investors. 

In our future research, we will employ this framework to analyze the relationships between buyers and sellers under different market conditions. Additionally, we expect that in the future research this framework will be exploited with non-financial applications, too, such as social networks\cite{scott2017social}, different communication channels\cite{onnela2007structure,newman2002email,isella2011s}, transportation\cite{guimera2005worldwide}, and co-authorship\cite{liu2005co} networks.
 

\section*{Methods}

\noindent \textbf{Dataset and categorization.} We use an investor-level transaction data set covering the period from 2004-01-01 to 2009-12-31 of all trades executed by domestic investors on the Helsinki Stock Exchange. The data set is composed of transactions belonging to $443\,556$ investors trading in 100 securities over 6 years. The analyzed security list includes the top 100 securities ranked by number of investors and transactions. Each investor in the data set is assigned to a sector group: Financial and Insurance, Government, Non-Financial, and Non-Profit companies, and Finnish Households. Households are further divided into five age groups: Under-Aged $(0,18]$, Young $(18,30]$, Middle-Aged $(30,50]$, Mature $(50,64]$, and Retired $(64,+\infty]$. Age attributes are derived for each transaction separately, taking into account the difference between the transaction date and the year of birth of the corresponding investor. All of these groups are also distributed geographically by assigning investor postal codes to 11 regions: Helsinki, Rest-Uusimaa, Eastern-Tavastia, South-West, Western-Tavastia, Central-Finland, South-East, Ostrobothnia, Northern-Savonia, Eastern-Finland, Northern-Finland. Together, these assignment rules form 99 investor categories. Each transaction in the dataset is assigned to one of these categories.

In terms of the number of investors, the largest categories are households in various regions and of different age. For example, the category ``Helsinki households, Middle-Age'' includes 58\,124 individual household investors. In contrast, there are 343 members in the category ``Helsinki financial-insurance companies''. The smallest categories are these ones for General Government, having 6-84 members, depending on the location. 

In terms of the average number of trading days, the most active investor category is middle-age individuals (households) in Helsinki area with approx. 1\,229 trading days/security over this data period. This is a high number, because this category trades a security {\em on average} in $\sim 80\%$ of all the unique trading days. Mature individuals (households), Financial and Insurance Institutions, and Non-Financial Companies, all in Helsinki area, followed with 1.097, 974, and 929 average trading days/securities, respectively. In contrast, general government investors categories located elsewhere than in Helsinki traded just on a few days over the whole period. On the other hand, if activity is measured by the number of transactions, then the most active category is ``Helsinki financial-insurance companies'' traded over 11 million times over the six-year period. Non-financial Companies and mature household investors, both in Helsinki, followed, having around 3.5 million and 2.3 transactions over the same period. Again, the categories for General Government located elsewhere than in Helsinki were inactive in terms of the number of transactions, having just a few transactions over the whole period.

The data that support the findings of this study are available from Euroclear Finland Ltd., however, are not available from the authors under the non-disclosure agreement signed with the data provider.
\\

\noindent \textbf{Transaction bootstrapping.} For network inference, we perform $B$ bootstrap iterations. For each bootstrap iteration, we uniformly re-sample with replacement the whole transaction data set under investigation, yielding a replica dataset of the same size as the original, but with randomly chosen transactions. More specifically, some transactions from the investigated dataset might appear more than once, while others might not be included at all, but the total number of transactions in the bootstrapped dataset will equal the number in the original dataset. Then, for each sampled transaction set $b$, we aggregate daily transaction records for each category, resulting in a net traded volume matrix $\mathcal{N}^b$, where $b\in\{1\ldots B\}$ and the columns of the matrix $w_{i}^{b}(t)$ are the net traded volume time-series of investor group $i$.
\\

\noindent \textbf{Mutual information (MI) estimation.} We estimate the MI values using net volume matrix. For simplicity, we assume that the joint distribution of net traded volumes is normal. Then we can calculate the MI analytically from Pearson’s correlation. If $w_j(t)$ are the net traded volume time-series in some security of investor group $j$, then the correlation coefficient between investor categories $i$ and $j$ is defined as $\rho_{ij}=\left[{\langle w_i(t) \times w_j(t)\rangle - \langle w_i(t)\rangle \times \langle w_j(t)\rangle}\right]/ \left[\sigma_i \times \sigma_j\right]$, where $\langle\bullet\rangle$ and $\sigma_\bullet$ denote the mean and standard deviation. Then MI is defined as 
$I(i,j)=-\frac{1}{2}\log\left(1-\rho_{ij}^2\right)$.
\\

\noindent \textbf{Network inference.} We apply a chosen network inference method to MI estimates obtained from the net traded volume matrix $\mathcal{N}^b$. A specific requirement for the inference method is that it is computationally efficient for handling a large bootstrap ensemble. For this reason, we have chosen the \emph{C3NET} \cite{altay2010inferring} inference method as our main network inference method, and MST as a comparison example. \emph{C3NET} is intended to infer a significant maximum MI network. This algorithm comprises three basic steps:
\begin{enumerate}
	\item MI values are estimated for each investor category pair.
	\item Each MI value estimate is tested against a null hypothesis of vanishing MI. The null-hypothesis $H_0: I(i,j)=0$, i.e. the MI between investor group $i$ and $j$ is zero.  
	
	In order to test the statistical significance of the MI estimates, we need to procure an appropriate null distribution. To do that, we independently resample with replacement dates, traded volumes, and categories completely eliminating any relationship between them. Then we aggregate daily transaction records for each category, resulting in a net traded volume matrix $\mathcal{\tilde{N}}^b$. We do this multiple times and each time we estimate MI values between pairs of investor groups. These values result in an estimate of the null distribution, which we use to find statistically significant MI values.

	\item From statistically significant MI values each investor group is allowed to keep a single link, with the strongest statistically significant MI value. The resulting binary network has at most $M$ relationships in a system of $M$ nodes.
\end{enumerate}
\emph{MST} is another strict network filtering algorithm. It keeps the subset of edges, that connect all nodes, while at the same time having the smallest possible total sum of edge weights. In order to use MST together with MI estimates, we transform MI values by multiplication with $-1$, then the network construction is straightforward:
\begin{enumerate}
	\item Sort pairwise MI estimates in a ascending order
	\item Take the node pair with the lowest $-1\times$ MI value and add a link between them in the network if it doesn't create a cycle in the network
	\item Remove the previously chosen MI value from the queue
	\item Repeat steps 2 and 3 until all nodes are connected and there are still unused MI values in the queue
\end{enumerate}

\noindent \textbf{Network aggregation.} The aggregation procedure takes an ensemble of $N$ independent undirected binary networks $\{G_k\}_{k=1}^N$ as an input and gives a single network $G$ as an output. Our procedure is methodologically similar with Ref.  \citen{de2012bagging}, where the inference was improved by aggregating bootstrap ensemble of gene regulatory networks, though we use the principle to aggregate layers, not bootstrap ensemble. In our demonstration the number of networks in the ensemble $N$ takes the values of $B$ when aggregating the bootstrap ensemble, $T$ when aggregating over non-overlapping time periods and $S$ when aggregating over securities. 

\begin{enumerate}
\item The network ensemble is aggregated into a weighted network $\{G_k\}_{k=1}^N \rightarrow G_w$, where the edge weights in the network $G_w$ correspond to the number of particular edge occurrences in the ensemble. For example, the weight of an edge between investor groups $i$ and $j$ is defined as 
\begin{equation}
n_{ij} = G_w(i,j) = \sum_{k=1}^N{G_k(i,j)},
\end{equation}
where $n_{ij}$ may assume integer values between $0$ and $N$.
\item We conduct a statistical hypothesis test to remove the need for an arbitrary link threshold parameter:\\
$H_0^{n_{ij}}$: The number of networks $n_{ij}$ in the ensemble with an edge between $i$ and $j$ is less than $n_0(\alpha)$, where $\alpha$ is the significance level.\\
We define $p$ as the probability of two investor groups being randomly connected. We estimate the probability $p$, for two groups to be connected by chance in an $N$ network ensemble, as the fraction of the actual number of edges in the ensemble $\sum_{i>j,k}\{G_k(i,j)\}$ to the number of all possible links in the ensemble $N\times (n(n-1)/2)$, where $n$ is the number of investor groups. Then $n_{ij}$ follows a binomial distribution, $\mathcal{B}(p, N)$ and 
\begin{equation}
p_{ij} = \mathbb{P}(n\geq n_{ij})=\sum_{n=n_{ij}}^N \binom{N}{n} p^n(1-p)^{N-n}
\end{equation}
is the probability of observing by chance the link between investor groups $i$ and $j$ more than $n_{ij}$ times.
\item In order to control the family-wise error rate, we leverage the strict Bonferroni multiple hypothesis test correction (MTC) procedure.  Following the Bonferroni procedure, we adjust the chosen significance level $\alpha$ by the number of tests ($n_{\text{tests}}$) we perform: $\alpha_{\text{adjusted}}=\alpha/n_{\text{tests}}$. Therefore, nodes in the aggregated network $G$ are connected if $p_{ij} < \alpha_{\text{adjusted}}$, where $\alpha_{\text{adjusted}}$ is the significance level.
\end{enumerate}

\noindent \textbf{Multilayer aggregation procedure.} 
\begin{enumerate}
\item For a set of securities $S$ and a number of non-overlapping inference periods $T$, we infer $S\times T$ networks $\{G_{st}\}_{(S\times T)}$. For all security$\times$inference period combinations, we perform transaction bootstrapping, network inference and bootstrap network ensemble aggregation, yielding the $\{G_{st}\}_{(S\times T)}$ ensemble. Each network represents significant relationships between investor groups for specific securities at  periods.
\item Then we apply the network aggregation procedure over securities for each period $t$, ${\forall t \in T:} \{G_{st}\}_{s=1}^S \xrightarrow{\text{network aggregation}} G_t$, we end up with an ensemble of networks $\{G_{t}\}_{t=1}^T$. Each of the $\{G_{t}\}$ networks represents significant relationships between investor groups that occur over multiple securities during period $t$. Similarly, if we apply the network aggregation procedure over time for each security $s$, $\forall s \in S:  \{G_{st}\}_{t=1}^T \xrightarrow{\text{network aggregation}} G_s$, we end up with an ensemble of networks $\{G_{s}\}_{s=1}^S$, where each of the networks $\{G_{s}\}$ represents the most important over time reoccurring relationships between investor groups in security $s$.

\item Finally, we aggregate the second layer of information. $\{G_{t}\}_{t=1}^T \xrightarrow{\text{network aggregation}} \widetilde{G} $ and $\{G_{s}\}_{s=1}^S \xrightarrow{\text{network aggregation}} \widehat{G}$ appropriately.  Both aggregation sequences lead to unique networks.
\begin{equation}
\begin{array}{l}
\{G_{st}\}_{(S\times T)} \quad \Longrightarrow \quad{\forall t \{G_{st}\}_{s=1}^S} \xrightarrow{\text{n. agg.}} {\{G_t\}}\quad
 \Longrightarrow \quad{\{G_{t}\}_{t=1}^T} \xrightarrow{\text{n. agg.}}{\widetilde{G}}
\\
 {\{G_{st}\}_{(S\times T)}} \quad \Longrightarrow \quad {\forall s \{G_{st}\}_{t=1}^T}\xrightarrow{\text{n. agg.}} {\{G_s\}}\quad
 \Longrightarrow \quad{\{G_{s}\}_{s=1}^S}  \xrightarrow{\text{n. agg.}} {\widehat{G}}
\end{array},
\qquad
\widehat{G} \neq \widetilde{G}
\end{equation}
 Both $\widetilde{G}$ and $\widehat{G}$ are accordingly equivalent to the illustrated networks (3) and (5) in Figure \ref{BaggingLayers}.
\end{enumerate}

\bibliography{sample,ref_list,bib_ref_books_editor}

\section*{Acknowledgements}
The research project leading to these results received funding from the EU Research and Innovation Programme Horizon 2020 under grant agreement No. 675044 (BigDataFinance).

\section*{Competing interests}
The authors declare no competing interests.

\section*{Contributions}
All authors designed the experiment, wrote and reviewed the main manuscript text. K.B. prepared all figures (figure 1 together with F.E.), and conducted the empirical analysis.

\end{document}